\documentclass{article}
\usepackage{spconf,amsmath,amssymb,graphicx}
\usepackage{booktabs,tabularx,array}
\usepackage{dblfloatfix}
\usepackage{float}
\usepackage{cite}
\usepackage{enumitem}
\usepackage[hidelinks]{hyperref}

\newcolumntype{Y}{>{\raggedright\arraybackslash}X}
\newcolumntype{C}{>{\centering\arraybackslash}X}
\setlist[enumerate]{leftmargin=1.4em,itemsep=1pt,topsep=2pt,parsep=0pt}

\makeatletter
\long\def\@makecaption#1#2{%
 \vskip 3pt
 \setbox\@tempboxa\hbox{#1. #2}
 \ifdim \wd\@tempboxa >\hsize #1. #2\par \else \hbox
to\hsize{\hfil\box\@tempboxa\hfil}
 \fi}
\makeatother

\newcommand{\DSB}{\Delta\mathrm{SIM}_{B}}
\newcommand{\DSA}{\Delta\mathrm{SIM}_{A}}
\newcommand{\psss}{\textsuperscript{***}}

\title{Decodability Is Not Controllability: Local and Anisotropic Speaker Control in Continuous-Latent TTS}

\name{Yuanzhong Zheng, Wei Huang, Ming Ma, Kailun Jin, Youhai Jiang\sthanks{Corresponding author.}}

\address{VoxNexus, Singapore}

\begin{document}
\ninept
\maketitle

\begin{abstract}
High linear decodability does not entail causal controllability in
continuous-latent text-to-speech (TTS). In VoxCPM2, position-matched probes and equal-norm
interventions across 30 speakers and 60 directed dyads expose this separation:
reference-audio states at the final residual acoustic language model layer
(RALM L7) retain 0.970 probe accuracy, yet aligned edits yield 0/60 target-dominant
dyads. Edits at early reference-audio positions produce the opposite pattern.
At L0, pair-aligned updates increase target similarity by 0.326 and achieve
target dominance in all 60 dyads, against at most 1/60 under any equal-norm
direction or position control. Independent speaker encoders, retrieval against
held-out speaker prototypes, and double-blind listening confirm that the
shifted voice is heard as the intended target. Cross-architecture
interventions in dots.tts distinguish backbone leverage from direct acoustic
control: backbone edits displace the source without target attraction,
while x-vector ablation reduces source similarity from 0.849 to 0.006.
These findings identify two requirements for training-free speaker control:
geometric specificity and a computational route that carries the intended
speaker condition into acoustic synthesis.
\end{abstract}


\section{Introduction}
\label{sec:intro}

Prompt-conditioned text-to-speech (TTS) systems synthesize target voices from
reference audio \cite{zhou2026voxcpm2,wang2023neural,le2023voicebox,casanova2022yourtts,du2024cosyvoice}.
While linear probes measure the speaker information
available in hidden states \cite{alain2016understanding,vanniekerk2021analyzing}, the critical distinction lies between reading
that information and using it to control the generated voice
\cite{hewitt2019designing,ravichander2021probing,elazar2021amnesic}. We expose
a direct counterexample to selecting control sites by probe accuracy: VoxCPM2's
L7 reference-audio states retain 0.970 accuracy yet yield 0/60 target-dominant
dyads under aligned intervention, whereas L0 aligned edits yield 60/60.
Locating speaker control therefore requires attending to both the direction
of an edit and its route into synthesis.

\begin{figure*}[!t]
\centering
\includegraphics[width=\linewidth,keepaspectratio]{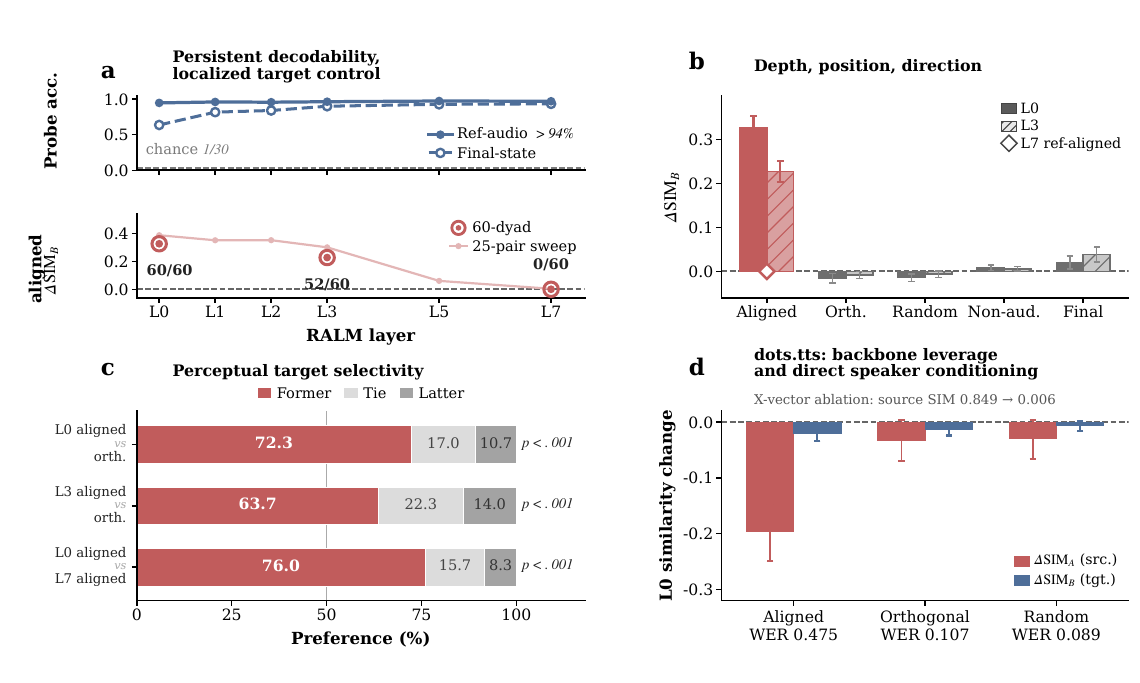}
\caption{(a) In VoxCPM2, speaker identity at reference-audio positions remains
highly decodable ($>$94\%, solid) while final-state decodability rises with depth
(dashed). Bottom: 60-dyad target gain at L0/L3/L7 (markers) with
target-dominant counts; a finer-grained six-layer sweep over 25 dyads (thin
line) resolves a sharp decline in target control at L5. (b) Early
reference-audio aligned updates produce the largest target gains among
equal-norm direction and position conditions. (c) Held-out-reference listeners
confirm perceptual direction and depth selectivity. (d) The dots.tts
interventions separate backbone leverage from direct speaker conditioning. At
L0, aligned updates produce the largest source displacement and highest WER;
mean target gain is non-positive across directions. X-vector ablation reduces
source similarity from 0.849 to 0.006. Uncertainty in (b,d) is evaluated by
dyad-bootstrap (95\% CIs; 60 directed dyads after averaging seeds and prompts
within each dyad).}
\label{fig:composite}
\end{figure*}

Activation patching, causal mediation, and representation steering test model
behavior under direct intervention \cite{meng2022locating,vig2020investigating,li2022emergent,turner2023steering,rimsky2024steering}.
In speech synthesis, recent work uses activation steering to manipulate
attributes such as emotion, style, and accent
\cite{xie2025emosteer,koriagin2026interpreting,wang2026geometric,yang2026activation}.
We couple position-matched probing with equal-norm geometric controls to
localize speaker control and distinguish target attraction from acoustic
disruption.

Target-specific control requires both increasing similarity to target speaker
$B$ and making the output closer to $B$ than to source speaker $A$.
An update can also suppress the source voice or disrupt speech without producing
the target voice. We distinguish \textbf{decodability}, the ability to predict
speaker identity from a state; \textbf{leverage}, the effect of changing that
state on speech; and \textbf{specificity}, whether the change steers speech
toward the intended target.

We study these properties in the residual acoustic language model (RALM) prefill
states of VoxCPM2 \cite{zhou2026voxcpm2}. Across 30 English LibriTTS-R speakers
\cite{koizumi2023librittsr} organized into 60 role-balanced directed dyads
(evaluated over two prompts and three generation seeds), we vary layer depth,
token position, and perturbation direction in an equal-norm factorial.
Independent speaker encoders, retrieval against held-out speaker prototypes,
and double-blind listening measure the identity of the generated voice.

First, the experiments establish a \textbf{sharp dissociation between decodability and target
control}: L0 reference-audio edits achieve target dominance in all 60 dyads,
whereas L7 edits yield 0/60 despite a probe accuracy of 0.970.
Second, control is \textbf{anisotropic and
position-specific}: pair-aligned updates at early reference-audio positions
produce target attraction, whereas equal-norm orthogonal updates reveal source-displacing
leverage, and final-state updates primarily affect intelligibility. The comparison
isolates the directions and positions that give leverage its target specificity.
Finally, \textbf{readout reach and direct conditioning identify the architectural routes
for speaker control}. In VoxCPM2, subsequent attention blocks carry early
reference-state edits to the synthesis readout. In dots.tts, backbone edits and
x-vector ablation distinguish behavioral leverage from a source-identity anchor
supplied directly to the acoustic head. These experiments identify the editable
states and conditioning inputs that a speaker-control interface must address.

\section{Method}
\label{sec:method}

\subsection{Model and probe}
\label{ssec:probe}

VoxCPM2 encodes reference waveforms into AudioVAE patches, maps them through a
local audio encoder, interleaves the resulting audio embeddings with text
embeddings, and processes the sequence through a TSLM and an eight-layer RALM;
a flow-matching LocDiT and the AudioVAE decoder then synthesize the waveform \cite{zhou2026voxcpm2}.
Interventions modify post-block RALM hidden states at prefill step 0 and retain
the intervened block's already-computed key/value (KV) cache. Subsequent
autoregressive steps use the standard forward computation. Early-block updates
propagate through later RALM blocks and their KV caches. The final RALM readout
selects the last sequence position to condition LocDiT. We define
\textbf{attention-mediated readout reach} as the capacity of a reference-audio
intervention to influence the final conditioning state through subsequent RALM
attention blocks. Early reference-audio edits have this route; post-block L7
reference-audio edits lie beyond the final mixing block.
Edits to the final sequence position at L7 reach the acoustic generator directly.

The matched probe contains 300 held-out utterances from the same 30-speaker
universe as the causal experiment. At layers L0, L1, L2, L3, L5, and L7, a
linear classifier predicts speaker identity from the reference-audio mean,
non-audio-token mean, final state, or whole-sequence mean. We use five-fold
cross-validation grouped by transcript identity and 1,000 full-label permutations with a
24-condition max-statistic family-wise correction. The probe matches the causal
experiment's speaker pool and reference/enrollment exclusions.

\subsection{Equal-norm activation steering}
\label{ssec:steer}

\begin{table*}[!t]
\caption{Speaker identity remains decodable across depth, while target
control concentrates at early reference-audio positions. Reference-audio probes
use the token mean. Target gain is $\DSB$, with dyad-bootstrap
95\% CIs; dominant counts are out of 60 directed dyads. Probe and intervention
samples are disjoint and share the same 30-speaker pool.}
\label{tab:matched}
\centering
\footnotesize
\setlength{\tabcolsep}{3.6pt}
\renewcommand{\arraystretch}{1.08}
\begin{tabularx}{\textwidth}{@{}lCCCCCC@{}}
\toprule
 & \multicolumn{3}{c}{\textbf{Reference-audio}} &
 \multicolumn{3}{c}{\textbf{Final-state}} \\
\cmidrule(lr){2-4}\cmidrule(lr){5-7}
\textbf{Layer} &
\textbf{Probe acc.} &
\textbf{Target gain [95\% CI]} &
\textbf{Dominant} &
\textbf{Probe acc.} &
\textbf{Target gain [95\% CI]} &
\textbf{Dominant} \\
\midrule
L0 & 0.947 & \textbf{+0.326 [0.298, 0.353]} & \textbf{60/60} & 0.637 & +0.020 [0.006, 0.035] & 1/60 \\
L3 & 0.963 & \textbf{+0.227 [0.204, 0.250]} & \textbf{52/60} & 0.900 & +0.038 [0.021, 0.055] & 4/60 \\
L7 & 0.970 & 0.000 [$-$0.002, 0.001] & 0/60 & 0.933 & +0.042 [0.027, 0.057] & 7/60 \\
\bottomrule
\end{tabularx}
\end{table*}

For source $A$, target $B$, shared synthesis text $x$, and RALM layer
$\ell$, let $h_A^\ell$ and $h_B^\ell$ be post-block prefill states. Before
subtraction, $h_B^\ell$ is length-aligned to the source sequence by linear
interpolation when shorter and truncation when longer. The pair-specific
direction is
\begin{equation}
d_{A\rightarrow B}^{\ell}=h_B^\ell-h_A^\ell .
\end{equation}
Let $M_p$ be a binary mask selecting position type $p$, broadcast across
channels, and let $z^\ell\sim\mathcal{N}(0,I)$ have the same shape as
$h_A^\ell$. For direction condition
$k\in\{\mathrm{align},\mathrm{orth},\mathrm{rand}\}$, define
$u_{p,\mathrm{align}}^\ell=M_pd_{A\rightarrow B}^\ell$,
$u_{p,\mathrm{rand}}^\ell=M_pz^\ell$, and
$u_{p,\mathrm{orth}}^\ell=M_pz^\ell-
\operatorname{proj}_{u_{p,\mathrm{align}}^\ell}(M_pz^\ell)$, where
$\operatorname{proj}_a(b)=\langle b,a\rangle_F a/\|a\|_F^2$. With
reference-audio aligned norm
$r^\ell=\|u_{\mathrm{ref},\mathrm{align}}^\ell\|_F$, every condition uses
\begin{equation}
\widetilde h_A^\ell=h_A^\ell+r^\ell
\frac{u_{p,k}^{\ell}}{\|u_{p,k}^{\ell}\|_F}.
\end{equation}
At L0, L3, and L7, the factorial tests reference-audio aligned,
reference-audio orthogonal, reference-audio random, non-audio aligned, and
final-state aligned conditions. The non-audio mask contains BOS, separator, and
text positions; the final-state mask contains only the last position. For each
dyad and prompt at a given layer, all five conditions share $r^\ell$, matching
total Frobenius norm across directions and positions.

Probe accuracy measures decodability; changes in source similarity and WER
measure behavioral leverage; target gain and target dominance measure
specificity. For generated speech $y$, the primary metric is matched-baseline
target gain
\begin{equation}
\DSB =
\mathrm{SIM}(y_{\mathrm{int}},B)-
\mathrm{SIM}(y_{\mathrm{base}},B),
\end{equation}
where SIM is cosine similarity between L2-normalized speaker embeddings. After
averaging seeds within prompts and the two prompts within each directed dyad, a
dyad satisfies the \textbf{target-dominant criterion} when both
$\DSB>0$ and
$\mathrm{SIM}(y_{\mathrm{int}},B)>\mathrm{SIM}(y_{\mathrm{int}},A)$.
The criterion combines target attraction with target-over-source preference in
the same dyad. The primary
factorial uses ECAPA-TDNN \cite{desplanques2020ecapa}. We score the same generated speech with
CAM++ \cite{wang2023campplus}, perform 30-way retrieval against five-utterance speaker prototypes
disjoint from all intervention references, and use Whisper WER \cite{radford2023robust} as an
intelligibility measure.

The directed dyad is the inferential unit. We report dyad-bootstrap 95\%
confidence intervals and two-way source/target cluster-robust 95\% confidence
intervals accounting for repeated speakers in both roles \cite{cameron2011robust}.

\section{Results}
\label{sec:results}

\subsection{Distributed speaker information, localized target control}
\label{ssec:decod}

The depth contrast separates speaker information from target control: L0 aligned
reference-audio edits achieve target dominance in 60/60 dyads, while L7 edits
yield 0/60 despite a probe accuracy of 0.970. Mean target gain falls from
$\DSB=+0.326$ at L0 to $0.000$ at L7; L3 retains a
gain of $+0.227$ with 52/60 target-dominant dyads (Table~\ref{tab:matched}). The final layer
thus retains decodable speaker information after reference-audio edits have lost
attention-mediated readout reach.

Speaker information is distributed more broadly than this control region.
Reference-audio probe accuracy stays within 0.947--0.973 across layers, while
non-audio accuracy rises from 0.223 at L0 to 0.903 at L7 and final-state accuracy
from 0.637 to 0.933. All reported probe conditions satisfy
$p_{\mathrm{FWE}}=0.001$ under the 24-condition permutation correction.
A finer-grained six-layer sweep over 25 dyads resolves the early control region
spanning L0--L3, followed by a sharp decline at L5 (Fig.~\ref{fig:composite}(a)); target dominance
reaches 0/25 at L5 and L7.

\subsection{Direction and position determine the steering response}
\label{ssec:conjunction}

Target specificity depends on both update direction and token position. At L0,
pair-aligned reference-audio steering exceeds equal-norm orthogonal and
random updates by $+0.342$ and $+0.340$ in target similarity, respectively
(Fig.~\ref{fig:composite}(b)). It also exceeds non-audio and final-state
aligned interventions by $+0.318$ and $+0.306$. All four 95\% confidence
intervals lie above zero under both dyad-bootstrap and two-way source/target
cluster-robust inference. At L3, aligned-minus-control gains range
from $+0.189$ to $+0.235$; at L7, aligned and direction-control contrasts
are centered near zero.

At L0, aligned reference-audio intervention both displaces
the source and attracts the target ($\DSA=-0.467$;
post-intervention $\mathrm{SIM}_B=0.467>\mathrm{SIM}_A=0.173$). Orthogonal
perturbations reduce source similarity ($\DSA=-0.195$) while
also reducing target similarity ($\DSB=-0.016$; 95\% CI
excludes zero). Orthogonal edits exert behavioral leverage by moving speech away
from both speakers; pair alignment channels that leverage toward the intended target.
Final-state interventions concentrate their leverage in intelligibility, raising
matched-baseline $\Delta\mathrm{WER}$ by $+0.113$ [$+0.058$, $+0.179$] at L0,
$+0.127$ [$+0.043$, $+0.251$] at L3, and $+0.096$ [$+0.053$, $+0.147$] at L7,
with low target-dominant counts (Table~\ref{tab:matched}). Reference-audio aligned intervention
achieves target gains with mean $\Delta\mathrm{WER}$ near zero (Table~\ref{tab:checks}).
Across the same 25 dyads used for the six-layer sweep, an L0
norm sweep at $0.5,1,2,5\times$ the native norm yields
$+0.186,+0.387,+0.302,+0.342$. The aligned direction supports target attraction
at all four scales, spanning a tenfold norm range, with the largest gain at the
native norm.

\subsection{Target specificity extends to held-out speaker references}
\label{ssec:generalize}

The target-speaker match extends across evaluators and reference utterances.
Independent CAM++ evaluation reproduces the depth ordering: aligned target gain
is $+0.237,+0.190,+0.001$ at L0/L3/L7, with
55/60, 43/60, and 0/60 target-dominant dyads (Table~\ref{tab:checks}). We also compare generated
speech with five-utterance prototypes disjoint from the intervention references.
At L0, ECAPA and CAM++ both retrieve the intended target as top-1 among 30
candidates for 46/60 dyads; at L3 this is 44/60 and 41/60, respectively, while at
L7 it falls to 0/60 for both. Successful retrieval against these independent,
multi-utterance prototypes establishes that target-speaker matching generalizes
beyond the intervention reference to the speaker level.

\begin{table}[ht]
\caption{Cross-encoder and held-out target evaluation. CAM++ gain denotes
aligned $\DSB$; dom.\ counts target-dominant dyads. Retrieval
is top-1 over 30 held-out speaker prototypes. CAM++ and retrieval use
the first prompt; Whisper $\Delta\mathrm{WER}$ uses both prompts. All counts are
out of 60 dyads; brackets report dyad-bootstrap 95\% CIs for
$\Delta\mathrm{WER}$.}
\label{tab:checks}
\centering
\footnotesize
\setlength{\tabcolsep}{2.8pt}
\renewcommand{\arraystretch}{1.08}
\begin{tabularx}{\columnwidth}{@{}lCCCC@{}}
\toprule
 & & \multicolumn{2}{c}{\textbf{Held-out target retrieval}} & \\
\cmidrule(lr){3-4}
\textbf{Layer} &
\textbf{CAM++ gain / dom.} &
\textbf{ECAPA} &
\textbf{CAM++} &
\textbf{Aligned $\Delta$WER} \\
\midrule
L0 & +0.237 / 55/60 & \textbf{46/60} & \textbf{46/60} &
$-0.003$ $[-0.018, 0.013]$ \\
L3 & +0.190 / 43/60 & 44/60 & 41/60 &
$-0.003$ $[-0.015, 0.007]$ \\
L7 & +0.001 / 0/60 & 0/60 & 0/60 &
$-0.001$ $[-0.003, 0.001]$ \\
\bottomrule
\end{tabularx}
\end{table}

\begin{table*}[!t]
\caption{Backbone leverage and target specificity in dots.tts under equal-norm
direction controls across 60 directed dyads,
two prompts, and three seeds. Similarity changes use ECAPA-TDNN and matched
baselines; WER is absolute Whisper word-error rate. Every condition has 0/60
target-dominant dyads. Aligned updates show the strongest source-displacement
and WER response at each layer; mean target gain is non-positive throughout.}
\label{tab:dots}
\centering
\footnotesize
\setlength{\tabcolsep}{10pt}
\renewcommand{\arraystretch}{1.08}
\begin{tabularx}{\textwidth}{@{}llCCC@{}}
\toprule
\textbf{Layer} &
\textbf{Direction} &
\textbf{Source change $\DSA$} &
\textbf{Target gain $\DSB$} &
\textbf{WER (absolute)} \\
\midrule
L0 & Aligned & $\mathbf{-0.197}$ & $-0.020$ & $\mathbf{0.475}$ \\
L0 & Orthogonal & $-0.032$ & $-0.014$ & 0.107 \\
L0 & Random & $-0.030$ & $-0.007$ & 0.089 \\
\midrule
L14 & Aligned & $\mathbf{-0.187}$ & $-0.004$ & $\mathbf{0.421}$ \\
L14 & Orthogonal & $-0.049$ & $-0.009$ & 0.095 \\
L14 & Random & $-0.022$ & $-0.010$ & 0.075 \\
\midrule
L27 & Aligned & $\mathbf{-0.046}$ & $-0.001$ & $\mathbf{0.231}$ \\
L27 & Orthogonal & $-0.014$ & $-0.002$ & 0.086 \\
L27 & Random & $-0.014$ & $-0.001$ & 0.068 \\
\bottomrule
\end{tabularx}
\end{table*}

\subsection{Direct acoustic conditioning anchors speaker identity in dots.tts}
\label{ssec:dots}

To trace control across distinct conditioning paths, we apply the
direction-control protocol to dots.tts \cite{lian2026dotstts}, a 2B-parameter continuous autoregressive TTS model
with a 28-layer backbone. Its acoustic head receives a global speaker x-vector
\cite{snyder2018xvectors} alongside backbone states. Speakers, dyads, prompts, seeds, and evaluators
are matched to the primary factorial. We
probe and intervene at backbone layers L0, L14, and L27 with aligned,
orthogonal, and random directions at reference positions.

X-vector ablation pinpoints a strong source-identity anchor at the direct acoustic
input: disabling it reduces mean source similarity from 0.849 to 0.006. The
x-vector and backbone states reach the acoustic head through separate routes;
backbone interventions edit one route while keeping the source-speaker x-vector
fixed.

The backbone itself carries both decodable identity and behavioral leverage.
Final-state probe accuracy increases with depth, from 0.260 at L0 to 0.437
at L14 and 0.530 at L27, measured by five-fold cross-validation. Early aligned
intervention is direction-specific (Table~\ref{tab:dots}):
$\DSA=-0.197$ at L0 versus $-0.032$ and $-0.030$ for
equal-norm orthogonal and random directions, with a weaker aligned effect at L27
($-0.046$). CAM++ reproduces the aligned source-displacement profile
($-0.154,-0.155,-0.023$ at L0/L14/L27). L0 aligned Whisper WER is 0.475,
versus 0.107 for orthogonal and 0.089 for random updates. Mean
$\DSB$ is non-positive in every condition, and all conditions
yield 0/60 target-dominant dyads. This contrast separates behavioral leverage
from target specificity within the same backbone. The acoustic head receives
edited backbone states alongside a source-speaker x-vector that remains fixed.
Backbone intervention and x-vector ablation therefore act on different
conditioning routes: one reveals effects on source similarity and intelligibility;
the other identifies the direct input that anchors speaker identity.

\begin{table}[H]
\caption{Double-blind listening over the 60 dyads, with five judgments
per dyad per comparison. (a) Target-speaker preference using held-out
references; ties remain in the denominator. Brackets show dyad-bootstrap 95\%
CIs; \psss$p<0.001$ (mixed-effects logistic regression). (b) Naturalness relative
to baseline on a $-3$-to-$+3$ CMOS scale, with 95\% CIs.}
\label{tab:human}
\centering
\footnotesize
\setlength{\tabcolsep}{4pt}
\renewcommand{\arraystretch}{1.08}
{\footnotesize
\textbf{(a) Target-speaker ABX preference}\\[2pt]
\begin{tabularx}{\columnwidth}{@{}Yc@{}}
\toprule
\textbf{Comparison (first vs.\ second)} &
\textbf{First-condition preference [95\% CI]} \\
\midrule
L0 aligned vs.\ L0 orthogonal &
\textbf{72.3\%} [67.2, 77.1]\% \psss \\
L3 aligned vs.\ L3 orthogonal &
\textbf{63.7\%} [58.4, 68.9]\% \psss \\
L0 aligned vs.\ L7 aligned &
\textbf{76.0\%} [71.1, 80.6]\% \psss \\
\bottomrule
\end{tabularx}

\vspace{4pt}
\textbf{(b) Comparative naturalness}\\[2pt]
\begin{tabularx}{\columnwidth}{@{}Yc@{}}
\toprule
\textbf{Comparison} &
\textbf{$\Delta\mathrm{CMOS}$ [95\% CI]} \\
\midrule
L0 aligned vs.\ baseline & $-0.12$ [$-0.21$, $-0.03$] \\
L3 aligned vs.\ baseline & $-0.07$ [$-0.15$, $+0.01$] \\
\bottomrule
\end{tabularx}
}
\end{table}

\raggedbottom
\subsection{Listeners hear the target-speaker shift}
\label{ssec:human}

Human judgments evaluate five comparisons over the 60 dyads.
Each comparison receives five double-blind judgments per dyad (300 per
comparison; 1,500 total), with candidate order randomized and counterbalanced.
ABX analysis uses all 300 trials per comparison in a binary mixed-effects
logistic regression (tested-condition preference = 1; alternative or tie = 0),
with random intercepts for dyads and listeners. Comparative naturalness uses a
seven-point CMOS scale from $-3$ to $+3$ (intervention minus matched baseline).

With an unseen target-speaker utterance as reference, listeners prefer L0 aligned
over its equal-norm orthogonal control in 72.3\% of trials, L3 aligned over its
orthogonal control in 63.7\%, and L0 aligned over L7 aligned in 76.0\%; all three
contrasts are significant at $p<0.001$ (Table~\ref{tab:human}). Listeners hear the target
specificity of early aligned steering while baseline naturalness is largely
preserved: mean $\Delta\mathrm{CMOS}$ is $-0.12$
(95\% CI $[-0.21,-0.03]$) at L0 and $-0.07$ ($[-0.15,+0.01]$) at L3.

\section{Discussion and Conclusion}
\label{sec:conclusion}

Speaker decodability and target control separate sharply under intervention:
states that decode speaker identity almost perfectly can lie beyond the mixing
path that reaches synthesis, while early reference-audio states admit precise,
target-directed edits. Equal-norm controls identify the
directions and positions responsible for this response. Held-out retrieval and
listening judgments connect the resulting target specificity to speaker identity
and perception.

The mechanism combines \textbf{geometric specificity with readout reach}. Subsequent
attention blocks carry early reference-audio edits to VoxCPM2's synthesis
readout; post-block L7 reference-audio edits lie beyond this mixing path.
The dots.tts comparison reveals a different routing arrangement: a direct
x-vector input anchors source identity alongside an editable backbone with
behavioral leverage. These findings give a concrete design principle for
training-free speaker control: expose editable states with downstream readout
reach, and coordinate their target-directed updates with any direct acoustic
speaker-conditioning inputs. Control-site selection must account for both
intervention geometry and the paths that carry speaker identity into synthesis.

% Official 2027 kit: optional fifth page may contain only references.
\clearpage
\bibliographystyle{IEEEbib}
\bibliography{refs-hca}

\end{document}